\documentclass[11pt]{article}
\usepackage{amsmath,amsfonts,amssymb,dcolumn,amsthm,euscript,bm}
\usepackage[english]{babel}
\usepackage{soul}
\usepackage[nosort]{cite}
\usepackage{filecontents}
\usepackage{color}
\usepackage[breaklinks=true]{hyperref}
\usepackage{mcite}

\numberwithin{equation}{section}
\hypersetup{colorlinks=true,
			urlcolor=blue,
			citecolor=magenta,
			linkcolor=blue,}

\def\be{\begin{equation}}
\def\ee{\end{equation}}
\def\beq{\be}
\def\eeq{\ee}
\def\bea{\begin{eqnarray}}
\def\eea{\end{eqnarray}}

\def\ol{\hat}
\def\bar{\overline}
\newcommand\bra{\langle\!\langle\,}
\newcommand\ket{\,\rangle\!\rangle}

\def\demi{\frac{1}{2}}
\newcommand\pa{\partial}
\def\d{{\mathrm d}}
\def\tpa{\tilde\pa_\tau} 
\def\k{D_\tau{\ln\sqrt g}}
\newcommand\Diff{\mathrm{Diff}_\Sig}
\newcommand\Weyl{\mathrm{Weyl}_{\cal M}}
\newcommand\Lie{\pounds}
\newcommand\gauge{\mathrm{gauge}}

\newcommand\Ph{\varPhi}
\newcommand\Ps{\varPsi}
\newcommand\Del{\varDelta}
\newcommand\Gam{\varGamma}
\newcommand\Sig{\mathsf{\Sigma}}
\newcommand\del{\delta}
\newcommand\lam{\lambda}
\def\fp{\varphi}
\def\A{{\pmb\gamma}}
\def\NF{{\cal N}}

\newcommand\nn{\nonumber}

\begin{document}

\title{\textbf{Weyl Symmetry in Stochastic Quantum Gravity\\}}
\author{\textbf{Laurent~Baulieu$^\dagger$, Luca~Ciambelli$^{\dagger\dagger}$
and Siye~Wu$^{\dagger\dagger\dagger}$}
\thanks{{\tt baulieu@lpthe.jussieu.fr},
\quad{\tt luca.ciambelli@polytechnique.edu},
\quad{\tt swu@math.nthu.edu.tw}}\\\\
\textit{$^\dagger$ LPTHE, Sorbonne Universit\'e, CNRS} \\
\textit{{} 4 Place Jussieu, 75005 Paris, France}\\
\textit{{$^{\dagger\dagger}$CPHT, CNRS, Ecole Polytechnique, IP Paris
}}\\ 
\textit{{Palaiseau, France}}
\\
\textit{{$^{\dagger\dagger\dagger}$Department of Mathematics,
National Tsing Hua University}}\\ 
\textit{{Hsinchu 30013, Taiwan}}}

\date{}
\maketitle

\begin{abstract}
We propose that the gauge principle of $d$-dimensional Euclidean quantum
gravity is Weyl invariance in its stochastic $(d+1)$-dimensional bulk.
Observables are defined as depending only on conformal classes of
$d$-dimensional metrics. We work with the second order stochastic quantization
of Einstein equations in a $(d+1)$-dimensional bulk.
There, the evolution is governed by the stochastic time, which foliates the
bulk into Euclidean $d$-dimensional leaves.
The internal metric of each leaf can be parametrized by its unimodular part
and conformal factor. 
Additional bulk metric components are the ADM stochastic lapse and a stochastic
shift.
The Langevin equation determines the acceleration of the leaf as the sum of
a quantum noise, a drift force proportional to Einstein equations
and a viscous first order force.
Using Weyl covariant decomposition, this Langevin equation
splits into irreducible stochastic equations, one for the unimodular part
of the metric and one for its conformal factor.
For the first order Langevin equation, the unphysical fields are the conformal
factor, which is a classical spectator, and the stochastic lapse and shift.
These fields can be gauge-fixed in a BRST invariant way in function of the
initial data of the process.
One gets observables that are covariant with respect to internal
reparametrization in each leaf, and invariant under arbitrary
reparametrization of the stochastic time.
The interpretation of physical observable at finite stochastic time is encoded
in a transitory $(d+1)$-dimensional phase where the Lorentz time cannot be
defined.
The latter emerges in the infinite stochastic time limit by an abrupt phase
transition from quantum to classical gravity.
\end{abstract}
\newpage

\section{Introduction}

It has been proposed in~\cite{Baulieu:2016qmc, *Baulieu:2018wju} that quantum
gravity might obey the laws of stochastic quantization~\cite{Parisi:1980ys},
governed by a second order rather than first order stochastic equation.
One theoretical motivation is the suppression of the problem of Lorentz time,
which one systematically encounters in the standard QFT formulations.
A physical motivation for using stochastic quantisation
in~\cite{Baulieu:2016qmc, *Baulieu:2018wju} was to predict that Lorentz time
can only emerge as a signature of the exit of inflation by an abrupt phase
transition from quantum to classical gravity, which gives also a heuristic
description to primordial cosmology as well as to the ultra-short distance
scattering of point-like particles. 

In this paper we refine the definition of the Langevin equation for gravity
of~\cite{Baulieu:2016qmc, *Baulieu:2018wju} making it genuinely geometric in
the stochastic bulk.
We arrive at the important conclusion on the definition of observables:
the statement that observables in $2$-dimensional gravity depend only on
the conformal classes of metrics~\cite{Polyakov:1981re, *Polyakov:1981rd}
can be extended to $d>2$.
We thus claim that the gauge symmetry principle of observables in
$d$-dimensional Euclidean gravity is the invariance under Weyl symmetry in
a $(d+1)$-dimensional bulk ${\cal M}$, modulo the internal reparametrization
of its Euclidean $d$-dimensional leaves $\Sig$.

Our proposal for the definition of quantum gravity is in fact inspired by the
observation that what really matters in classical gravity is the propagation
of conformal classes of spatial metrics. 
To the best of our knowledge, this property was firstly advocated in the
physical literature in \cite{York72, *York73}, where some of the Einstein
equations of motions were cornered out as dictating only physically
irrelevant propagation of constraints.
See also the more recent works \cite{Anderson:2004bq, *Gomes:2010fh}, where
the ``relativity of local size'' is implemented to find solutions to the
Einstein equations and equations in York's conformal technique to solve the
initial value problem.
This implies that the initial physical data for solving Einstein equations of
motion only concern conformal classes of spatial metrics, and our definition
of observables of quantum gravity implies this classical property.\footnote{In
fact, mathematicians found already in $1925$ the relevance of Weyl symmetry
for solving Einstein equations \cite{Thomas199, *Thomas2, *Thomas3}.} 
It will indeed appear that the conformal factor is left invariant in the last
steps of the stochastic time evolution where it is not submitted to relevant
quantum effects.

In practice, we propose that in quantum gravity observables must be
reparametrization covariant functionals of the unimodular part of the metric,
that is, the physics of quantum gravity is carried by conformal classes of
metrics. This is motivated by the work of York \cite{York72,*York73}, which demonstrated the role of conformal classes of spatial metrics in the description of the evolution in classical gravity and the observables. We show in our quantization procedure that functionals of
the unimodular part of the metric possess emblematic properties at late stochastic time.

We emphasize that  defining  unimodular  classical gravity has been one of the  early ideas of Einstein, in view of the existence a non-vanishing cosmological constant. Papers have investigated the notion of     unimodular  gravity,  as e.g.  \cite{cicci} and enclosed references.  
 There are basically  two standard  formulation of unimodular gravity: one  imposes $\det(g_{\mu\nu})=1$ as a gauge choice while the other imposes it as a constraint.  Our point of view is that in a formal perturbative treatment of  quantum  gravity using the Einstein action as a classical action plus BRST invariant gauge-fixing term, one can  do a BRST invariant gauge fixing of the metrics $g_{\mu\nu}$
 with   gauge functions $\sqrt { \det(g_{\mu\nu}})-1$ and  $\pa_\nu \hat  g^{\mu\nu}$, where 
 $\hat  g_{\mu\nu}$ is the unimodular part of the metrics, which gives a gauge-fixed BRST  local action that defines a  perturbation expansion for the correlators of $\hat  g_{\mu\nu}$, while $\sqrt { \det(g_{\mu\nu}})=1$.  In this way any given classical solution of Einstein equations  can be rewritten as a unimodular metric  by an appropriate choice of coordinates system. We can then extend it  at the   perturbative quantum level using a unimodular propagating metric, while the conformal factor is spectator.
 This is compatible with the conclusion  that we   draw in this paper by studying the stochastic quantization of gravity.

We postulate the existence of the stochastic bulk, whose leaves host a
Euclidean $d$-dimensional theory.
The quantum stochastic bulk correlators asymptote to the ones of classical
Euclidean gravity at infinite stochastic time.
This is of course a completely different framework than the classical
approach of \cite{York72, *York73}, where it was showed that it is the
$d$-dimensional Lorentz spacetime itself that can be foliated by spatial
$(d-1)$-dimensional leaves, and only conformal classes of spatial metrics
matter when solving the classical Einstein evolution.
 
It must be clear also that the theory is not conformal gravity at late
stochastic time, since the stochastic evolution is based on Einstein
equations of motion, which are not Weyl covariant.
Having equations of motion that are not scale invariant is actually not in
contradiction with the postulate that quantum observables of quantum gravity
are Weyl invariant functionals of the metric.
In fact this definition of observables does not imply that the dynamical
evolution in the stochastic bulk is Weyl invariant.

Because we propose that the gauge principle for quantum gravity observables
is that of Weyl symmetry, a BRST invariant gauge-fixing of the unphysical
fields will be needed.
By using an ADM type parametrization for the stochastic bulk
\cite{Arnowitt:1962hi}, the unphysical geometrical fields will be identified
as the conformal factor of the $d$-dimensional metric (in the first order
Langevin theory) and the lapse and shift functions of the $(d+1)$-dimensional
stochastic bulk.
The gauge-fixing procedure will be implemented in the $(d+1)$-dimensional
quantum field theory defined by stochastic quantization, namely within the
context of equivariant topological quantum field theory.	 

Contrarily to standard $d$-dimensional quantization methods, stochastic
quantization possesses the ingredients to clarify under which circumstances
can one get an ``emerging" Lorentz time, by checking (a hard task) the
possibility of an analytic continuation of the stochastic time correlators
of the unimodular metric on each Euclidean $d$-dimensional leaf, computed
at a finite stochastic time. 

Let us stress that having an acceleration term in the stochastic time
evolution equation of the metric, heuristically motivated
in~\cite{Baulieu:2016qmc, *Baulieu:2018wju}, is a new input which in
cosmological models could explain the exit from inflation by a sharp
transition and the emerging of the classical Lorentz time.
Such an acceleration tensor along the normal of each leaf has not been often
used in the geometry of foliated spaces and its investigation is quite
inspiring for visualizing the dynamics of leaves.

This work makes precise the idea that there is no limit of infinite stochastic
time in the quantum phase of gravity except for $\hbar=0$,
  modulo some formal perturbations
for taking into account the possibility of emitting and absorbing perturbative
traceless gravitons (namely, excitations of the unimodular part of  the classical backround $\hat g_{\mu\nu}$).
Such perturbative modifications  are in fact compatible with the framework of
stochastic quantization.
In fact, instead of having a hypothetical $d$-dimensional equilibrium
distribution for correlators of the metric in the $\tau=\infty$ limit and
$\hbar\neq0$, there are oscillations in the stochastic time because of the
acceleration term. These oscillations express the dynamics of quantum gravity.
They may sharply stop by a brutal transition toward classical gravity where,
effectively, the limit at infinite stochastic time can be reached, and the
whole theory can be directly computed using standard Euclidean QFT methods in
$d$ dimensions.
The effects of quantum gravity can thus only manifest themselves at finite
stochastic time, within the framework of a $(d+1)$-dimensional quantum field
theory, giving a specific ultra-short distance physics
\cite{Baulieu:2016qmc, *Baulieu:2018wju}. 

A non-trivial part of the program for building covariant Langevin equations
of gravity relies on the equivariant topological supersymmetry hidden in all
Langevin equations~\cite{Parisi:1980ys, 1983PhLB..129..432G}.
Indeed, in the case of stochastic quantization of a system with local
symmetries, one needs additional gauge degrees of freedom in the stochastic
bulk \cite{Baulieu:1988jw, *Baulieu:1989qh, *Baulieu:1993ff,
Baulieu:2000xi}.\footnote{In Appendix \ref{App1}, we sketch for the sake of
completeness the improvements needed to define stochastic quantization of a
theory with gauge invariance and impose a gauge restoring force along its
gauge orbits.}

In this paper the equivariance is with respect to $\Diff\ltimes\Weyl$,
namely the semi-direct product of the $d$-dimensional diffeomorphism symmetry
in each leaf and the Weyl symmetry in the whole bulk.
In fact $\cal M$ is asssumed to be $\Sig\times\mathbb{R}$ and $\Diff$ acts on
$\Sig$ and hence on $\cal M$ and on the space of metrics on $\cal M$.
This equivalence allows a proper definition of observables, in the same spirit
of the definition of topological observables in a topological quantum field
theory.
We remark that although the product structure $\Sig\times\mathbb{R}$ of
$\cal M$ seem to prevent topology changes of the leaf $\Sig$ in the
$\tau$-evolution, we will postulate that the metric on $\cal M$ satisfies
a highly non-linear second order Langevin equation \eqref{Lan}.
Solutions to such equations typically develop singularities in the course of
time evolution, even though the initial data are completely smooth.
Some of these singularities can be interpreted as topology changes of the
leaf $\Sig$ in the $\tau$-evolution.
However, we will be mostly concerned with the smooth solutions without
topology change in this paper.

To achieve our program, some difficulties have to be overcome by decomposing
irreducibly all quantities under the representation of $\Diff\ltimes\Weyl$.
This is in fact a necessary task to get the correct covariant expression of
the stochastic time acceleration of the metric.
For this, we have used the ADM decomposition, substituting basically the
Lorentz time of the current ADM formalism \cite{Arnowitt:1962hi} with the
stochastic time, with a different dynamics led by $d$-dimensional Einstein
equations of motions plus additional forces to ensure well-defined drift
forces along the gauge orbit directions.
To unravel the Weyl covariance and implement the unimodular decomposition
we found the recent work of \cite{Kiefer:2017nmo} extremely useful, in the
spirit of \cite{Thomas199, *Thomas2, *Thomas3}. 
More recent works have dealt with Weyl symmetry, in a similar
fashion as our here, in various disparate domains, such as scale-invariant
gravity \cite{Anderson:2002ey}, holography \cite{Ciambelli:2019bzz} and
hydrodynamics \cite{Diles:2017azi}.

The unimodular decomposition implemented in this paper will be dictated by 
a change of field variables which is pivotal to our results
expressing observables as Weyl invariant functionals.
We call this method the ``golden rule'', which allows us to decompose neatly
all geometrical quantities in function of Weyl invariant $\Sig$-tensors plus
terms depending on the conformal factor.
In fact, our ``golden rule'' bears some technical and conceptual similarities
with the ``dressing field method'' used in \cite{Francois:2018kcr}.

Compared with a vast literature on quantum gravity, the novelty in our work
is to postulate that time is not a fundamental parameter to order phenomena.
Rather it may (or may not) emerge, depending on the configuration of the
stochastic leaves at a given value of the stochastic time.
Moreover, we give a physical meaning to the so-called stochastic time.
We claim that the latter is a physical microscopic time that is an alternative
to the Lorentz time, when the latter does not exist.

Prior to establishing a rigorous connection between the Dirac observables,
say in the Wheeler-de Witt quantization, and the observables that we introduce,
we point out a substantial conceptual difference between our approach and the
traditional methodology.
In the former, one first defines a Euclidean quantum field theory using a
more general stochastic quantization process and ask afterward the physical
interpretation of the correlators of the metrics in each constant stochastic
time leaf.
In the latter, one relies on the validity at the quantum level of the
classical notions such as Dirac observables, as in \cite{Crnkovic:1986ex}.
Often, the presentation does not resolve key problems such as constraints
posed by Wheeler-de Witt or the indefiniteness of path integrals in gravity.

In cases where the Wheeler-de Witt superspace technique does lead to well
defined computations in quantum gravity, we can reproduce these results by
the stochastic approach because they are obtained from a well defined
$4$-dimensional path integral.
Some perturbative aspects of quantum gravity enters the discussions of Dirac
observables, as in \cite{Marolf:2009wp}, but they can also be accommodated
within the context of stochastic quantization, which is generally well defined
at infinite stochastic time, order by order in perturbation theory, using
relevant cutoffs.
In cases where statistical physics is described directly in term of a Boltzmann
distribution, Langevin-equation techniques are not needed, and the situation
in Euclidean quantum field theory is completely analogous.
Notice, however, that for Dirac observables, the gauge symmetry comes only
from diffeomorphisms, but the superspace analysis could perhaps be enriched
by the use of Weyl invariance.
Indeed, this invariance plays a key role in our definition of the observables,
as well as in the work of York. 

This article is structured as follows.
Section~\ref{sec2} introduces the technical structure to describe the
stochastic bulk: its ADM decomposition and the speed and acceleration of the
leaves.
Section~\ref{sec3} is devoted to the Langevin equation and its covariance on
the stochastic leaves. We then analyze various contributions from the
Langevin equation, both physically and mathematically.
To enlighten the Weyl properties of the Langevin equation, Section~\ref{sec4}
deals with the decompositions into the traceless and trace parts of various
quantities.
The ``golden rule'' described above is made explicit in Section~\ref{sec5}
to perform the unimodular decomposition of the traceless and trace parts of
the Langevin equations.
Eventually, Section \ref{sec6} discusses the observables in the quantum
stochastic bulk.
There we show the presence of a residual $x$-independent Weyl symmetry which
is used, together with the BRST symmetry, to localize the conformal factor.
We then specialize to some relevant sub-cases of the Langevin equation.
This is possible because we use a second order Langevin equation which involves
a physical dimensionful parameter $\Delta T$, whose value defines the relative
strength between the acceleration and viscous effects in the second order
Langevin equation. 

After concluding remarks in Section~\ref{sec7}, Appendix~\ref{App1} sketches
a proof of the invariance of the gauge-invariant-observables evolution under
the addition of a gauge fixing restoring term in the first order Langevin
equation.

\section{ADM decomposition of the stochastic bulk}\label{sec2}

To put in a geometrical framework the heuristic second order Langevin equation
depicted in \cite{Baulieu:2016qmc, *Baulieu:2018wju}, the language of
foliation inside the stochastic bulk is most useful.

If we call $x^\mu$ the coordinates of classical Euclidean Einstein
$d$-dimensional gravity with metric $g_{\mu\nu}(x)$ and action
$S=\int\d^d x\sqrt g R(g_{\mu\nu})(x)$ (in suitable units), the basic idea
of stochastic quantization is the extension 
\bea
x^\mu &\to & x^A\equiv(x^\mu,\tau),\nn\\
g^{\mu\nu}(x) & \to &
(g^{\mu\nu}(x,\tau),\ g^{\mu\tau}(x),\ g^{\tau\tau}(x,\tau)).
\eea
The aim is to define a quantum field theory in the $\{x,\tau\}$ space, with
a flow of correlators of the $x$- and $\tau$-dependent fields, toward a
certain limit when $\tau\to\infty$.\footnote{For renormalizable theories
such as Yang-Mill, $\tau$-dependent correlators flow smoothly when
$\tau\to\infty$ toward the correlators of the standard path integral in
$d$-dimensions, but in gravity the situation is different because there is
no equilibrium distribution for $\hbar\neq 0$ and there is no smooth limit
to classical gravity.}

To describe this geometrical system, it is appropriate to use the ADM
parametrization \cite{Arnowitt:1962hi} for the pseudo-Euclidean squared length
\beq\label{ADM}
\d s^2=-N^2\d\tau^2+(\d x^\mu+N^\mu\d\tau)g_{\mu\nu} (\d x^\nu+N^\nu\d\tau)
\eeq
of any given infinitesimal line element in the stochastic bulk.\footnote{The
signature of the total $(d+1)$-dimensional metric is $(-,+,\cdots,+)$.}
In this expression $N^\mu=N^2g^{\mu\tau}$ is the stochastic shift vector and
$N$ is the lapse, such that $g_{\tau\tau}=-N^2+N_\mu N^{\mu}$.
The determinant of the $(d+1)$-dimensional metric is $-N^2g$, where
$g=\det(g_{\mu\nu})>0$.

There is no demand for full reparametrization invariance in the total space
${\cal M}=\{(x^\mu,\tau)\}$.
Rather, the pseudo-Euclidean $(d+1)$-dimensional bulk ${\cal M}$ is foliated
by the stochastic time $\tau$, with equal-stochastic-time Euclidean
$d$-dimensional leaves $\Sig\equiv\Sig(x^\mu,\tau)$ having internal Euclidean
metric $g_{\mu\nu}(x,\tau)$.
Upper indexes $\mu,\nu\ldots$ are lowered by the $d$-dimensional tensor
$g_{\mu\nu}(x,\tau)$ in a way that preserves the reparametrization symmetry
of $\Sig$.
The foliation using the coordinate $\tau$, gives an absolute meaning to
$\tau$, modulo some possible one-dimensional reparametrization
$\tau\to\tau=\tau'(\tau)$.

For this reason, we can postulate that $N^\mu=N^2g^{\mu\tau}=N^\mu(x)$ is
$\tau$-independent.
This condition is preserved by diffeomorphisms on $\Sig$ of the form
$x^\mu\to x'^\mu(x)$, whose infinitesimal transformations are represented
by the Lie derivative $\Lie^{\Sigma}_\xi$ along a $\tau$-independent vector field
$\xi^\mu(x)$ (the operation $\Lie^\Sigma_\xi$ only involves $g_{\mu\nu}$).
Note however that $N_{\mu}=g_{\mu\tau}$ is $x$- and $\tau$-dependent, as well as
$N(x,\tau)$.

This decomposition is for a different purpose from that of
\cite{Arnowitt:1962hi}, where the Minkowski time was used to define the
foliation.
Here the foliation parameter is the stochastic time.
The physical meaning differs completely.

We will shortly consider Weyl covariance in our presentation: it dilates
locally the metric fields but not the coordinates $x^\mu$ and $\tau$.
Eq.~\eqref{ADM} shows that $N^\mu$ is Weyl invariant from the beginning,
since $\d x^\mu+N^\mu d\tau$ must be a Weyl invariant one-form.
This condition is the tip of an iceberg, made of all Weyl transformations of
fields and curvatures in ${\cal M}$. The technology is as in, e.g.,
\cite{Kiefer:2017nmo}, where it is actually used in the different context
of standard Hamiltonian classical gravity.

Symmetrization is defined as $T_{(\mu\nu)}=\demi(T_{\mu\nu}+T_{\nu\mu})$ and
$\nabla_\mu$ is the Levi-Civita connection of $\Sig$ with Christoffel
coefficients $\Gam^\gamma_{\alpha\beta}(g_{\mu\nu}(x,\tau))$ (involving no
$\tau$ derivatives).
The Riemann tensor of $\Sig$, $R^\alpha{}_{\beta\gamma\delta}(g_{\mu\nu})$,
is derived from the Christoffel symbols by the standard formula, except that
$g_{\mu\nu}$ is function of $x^\mu$ and $\tau$.
We will now define the speed and acceleration of $\Sig$ along its normal
vector. 

\subsection*{Speed and acceleration of a leaf}

Each leaf $\Sig$ in the foliation is defined at a fixed value of stochastic
time $\tau$, has internal metric $g_{\mu\nu}(x,\tau)$ and normal vector in
the stochastic bulk $\bm{N}$ with corresponding one-form $\bm{n}$:
\beq
{\bm{N}}\equiv N^A\pa_A=\frac{1}{N}\pa_\tau-\frac{N^\mu}{N}\pa_\mu, \qquad
\bm{n}\equiv N_A\d x^A=-N\d\tau,
\eeq
where we have normalized it as $N_AN^A=-1$.

Following e.g. \cite{Gourgoulhon:2005ng, *Gourgoulhon:2007ue}, we can define
the projector onto $\Sig$ as
\beq
P^A{}_B\equiv \delta^A{}_B+N^A N_B,\quad
P^A{}_\mu g_{A\nu}=g_{\mu\nu},\quad
P^A{}_B N_A=0.
\eeq
This projector allows to extend objects defined on the hypersurface to the
full stochastic bulk. 

The extrinsic curvature (second fundamental form) of a leaf in $\Sig$
represents the variation of the internal metric along the
hypersurface-orthogonal direction $\bm{N}$.
Hence
\beq\label{Extri}
K_{\mu\nu}\equiv\demi\Lie_{\bm{N}}\,g_{\mu\nu}=
\demi\big(N^A\pa_Ag_{\mu\nu}+\pa_\mu N^Ag_{A\nu}+\pa_\nu N^Ag_{\mu A}\big),
\eeq
which gives the explicit result
\beq
K_{\mu\nu}=\frac{1}{2N}
\big(\pa_\tau g_{\mu\nu}-N^\alpha\pa_\alpha g_{\mu\nu}-\pa_\mu N^\alpha
g_{\alpha\nu}-\pa_\nu N^\alpha g_{\mu\alpha}\big)=
\frac{1}{2N}(\pa_\tau g_{\mu\nu}-\nabla_\mu g_{\nu\tau}
-\nabla_\nu g_{\mu\tau}).
\eeq 
Its trace is
\beq\label{traceK}
K\equiv g^{\mu\nu}K_{\mu\nu}=\frac1N(\pa_\tau\ln\sqrt{g}-\nabla_\mu N^\mu).
\eeq
 
The extrinsic curvature can be extended in ${\cal M}$ using the projector
$P^A{}_B$,
\beq
K_{AB}\equiv P^\mu{}_A P^\nu{}_B K_{\mu\nu}.
\eeq
One can verify
\beq\label{Kext}
K_{\tau\tau}=N^\mu N^\nu K_{\mu\nu},\qquad
K_{\tau\mu}=N^\alpha K_{\alpha\mu}=N^\alpha K_{\mu\alpha}=K_{\mu\tau}.
\eeq

It is convenient to define the $N$-independent stochastic speed 
$D_\tau g _{\mu\nu}$ of a leaf along its normal~as
\beq\label{speed}
D_\tau g_{\mu\nu}\equiv 2NK_{\mu\nu}=
\pa_\tau g_{\mu\nu}-\nabla_\mu g_{\nu\tau}-\nabla_\nu g_{\mu\tau}
\eeq
and introduce the rate of evolution of this speed, which we call the
acceleration $\A_{\mu\nu}$ of the leaf along its normal $\bm{N}$
\be\label{acce}
\A_{\mu\nu}\equiv N\Lie_{\bm{N}}(D_\tau g _{\mu\nu})
=(\pa_\tau-N^\alpha\pa_\alpha)D_\tau g_{\mu\nu}
-2D_\tau g_{\alpha(\mu}\pa_{\nu)}N^\alpha.
\ee
The acceleration $\A_{\mu\nu}$ is the specific part of the $(d+1)$-dimensional
Riemann tensor $R^A{}_{BCD}$ that is covariant in the leaf at constant $\tau$
and contains the term $\pa_\tau^2g_{\mu\nu}$ but no derivative of the lapse
function $N$.

Both the speed $D_\tau g _{\mu\nu}$ and acceleration ${\A}_{\mu\nu}$ are
covariant tensors in the leaf with respect to diffeomorphisms with
$\tau$-independent parameters $\xi^\mu(x)$.
Moreover $N$ and $N^\mu$ are respectively a scalar and a vector for such
diffeomorphisms, denoted from now on as $\Diff$.

Let us stress again that both $D_\tau g _{\mu\nu}$ and $\A_{\mu\nu}$ are
constructed to be independent on $N$.
This plays an important role for understanding the stochastic evolution of
the leaves.

\section{The leaf-covariant Langevin equation}\label{sec3}

The necessity of an acceleration term in the stochastic evolution of the
metric $g_{\mu\nu}(x,\tau)$ was suggested in
\cite{Baulieu:2016qmc, *Baulieu:2018wju} on the basis of physical arguments,
so that the Langevin equation equates to a combination of a drift force
proportional to Einstein equations of motion, a viscous force containing
$\pa_\tau g_{\mu\nu}$ and a noise $\eta_{\mu\nu}$ multiplied by the square
root of the Planck constant $\sqrt\hbar$.
In fact, \cite{Baulieu:2016qmc, *Baulieu:2018wju} discusses the consequences
for primordial cosmology and the short distance behavior of particles of such
a second order Langevin equation for the stochastic quantization of gravity.
We will make this equation $\Diff$-covariant.

Beyond specific arguments for Gravity, one may generically justify the need
of a stochastic acceleration term as follows: in the stochastic evolution of
a massive particle with drift force $U'(x)$, the Langevin theory considers
Newton laws of mechanics for a large number of particles with this conserved
drift force plus some uncertainties and loss of information on the details of
the evolution.
The original Langevin equation was actually a second order one
\[ m\ddot x=-U'(x)-\alpha\dot x+\beta\eta. \]
It is often a hard task to prove that one can neglect the ``inertial'' term
$m\ddot x$, getting the simplified Langevin equation
\[ \alpha \dot x=-U'(x)+\beta\eta, \]
where $\beta$ can be eventually related to the statistical temperature of
the system.
One must prove case by case that the acceleration term can be neglected when
approaching an equilibrium.
The existence of an equilibrium itself must be demonstrated, depending on
the chosen potential $U$. 
The possibility of a phase transition is a delicate question. 
Appendix \ref{App1} sketches how to generalize the Langevin equation when
there is gauge symmetry, by defining at the same time both the stochastic
evolution of gauge degrees of freedom and their stochastic gauge-fixing.

For the stochastic quantization of gravity it was suggested in
\cite{Baulieu:2016qmc, *Baulieu:2018wju} that the acceleration term of the
Langevin equation is essential and cannot be neglected since one cannot
softly approach the limit $\tau\to\infty$ because otherwise there would be a
Euclidean equilibrium distribution in quantum gravity, which is not the case.

We must address the geometric aspects of the second-order Langevin equation
of gravity by giving a leaf-covariant formulation of it, with the
understanding that the leaf is the analog of a particle with internal
structure $g_{\mu\nu}$ and trajectory parametrized by $\tau$. 
The $\Diff$-covariance of the stochastic process will be obtained by giving
a central role to the leaf-tensors $D_\tau g_{\mu\nu}$ and $\A_{\mu\nu}$.

For this purpose, we postulate that the second-order Langevin equation is
\beq\label{Lan}
{\Del T {\A}_{\mu\nu}=-\NF G_{\mu\nu \rho\sigma}
\frac{\delta S}{\delta g_{\rho\sigma}}-D_\tau g_{\mu\nu}+2Ng_{\mu\nu}
+\sqrt\hbar\,\eta_{\mu\nu},}
\eeq 
where $\Del T$ is the dimensionful constant introduced in
\cite{Baulieu:2016qmc, *Baulieu:2018wju}.
 
The $\Sig$-tensor $G_{\mu\nu \rho\sigma}$ is a function of $g_{\mu\nu}$ that
will be defined in Eq.~\eqref{ker}, while
$\NF G_{\mu\nu\rho\sigma}(g_{\alpha\beta})$ is a kernel that factors
Einstein equations of motion. A discussion of its decomposition in trace
and traceless parts will shortly follow. 
 
In order to get a scalar in each leaf with the right conformal weight,
the factor $\NF $ must be equal to 
\bea\label{facs}
\NF\equiv \alpha N_\mu N^\mu+\beta N^2,
\eea
where $\alpha$ and $\beta$ are numbers.
To have an equilibrium distribution at $\tau=\infty$ independent of the
choice of the kernel $\NF G_{\mu\nu\rho\sigma}$, the noise distribution is
related to the kernel in the following way\footnote{The theorem about the
kernel independence property for observables, and thus of the arbitrariness
of the parameters needed in its expression, is a general property
\cite{Parisi:1980ys}, and can be proven e.g. by transforming the Langevin
equation in a Fokker-Planck equation. See Appendix~\ref{App1}.}
\beq\label{noise}
\bra{\cal F}[\eta_{\mu\nu}]\ket^\tau\equiv\int[\d\eta _{\mu\nu}]\,
{\cal F}[\eta_{\mu\nu}]\,\exp\Big[\!-\demi\int\d^dx\d\tau\sqrt g\,
\eta_{\alpha\beta}\NF^{-1}\,G^{-1\alpha\beta\rho\sigma}\eta_{\rho\sigma}\Big].
\eeq
We will use later the freedom in the choice of the coefficients $\alpha$ and
$\beta $ and choose
\bea\label{conds}
\alpha =1,\qquad\beta=0
\eea
that will facilitate examining the properties of the stochastic process for
observables which we will define by demanding Weyl invariance.

The principle of stochastic quantization is that correlation functions of the
noise are defined as an input, \cite{Parisi:1980ys}, given by \eqref{noise}.
Correlation functions of the fields are then computable at all possible values
of the stochastic time $\tau$ because
$g_{\mu\nu}=g_{\mu\nu}(\eta_{\alpha\beta})$ is a composite function of $\tau$
if $\eta_{\alpha\beta}$ solves the Langevin equation \eqref{Lan} with suitable
initial conditions.

Indeed, with suitable initial conditions at an arbitrarily chosen initial
value of the stochastic time, the differential equation \eqref{Lan}
determines $g_{\mu\nu}(x,\tau)$ as a function
$g_{\mu\nu}[\eta_{\alpha\beta}(x,\tau)]$ of the noise $\eta_{\mu\nu}(x,\tau)$.
Notice that for $\hbar=0$, the Langevin equation is nothing but a flow
equation toward the solutions of classical equations of motion, in which case
the correlations functions are just exactly centered on the solution of the
flow equation. 
Since all correlation functions of the noise $\eta_{\mu\nu}(x,\tau)$ are
computable by Eq.~\eqref{noise} after inserting the solution of the Langevin
equation in any given functional ${\cal O}[g_{\mu\nu}]$, if one substitutes
the correlation functions $ {\cal O}[g_{\mu\nu}](\eta)(x,\tau)$ in place of
$\cal F$ in Eq.~\eqref{noise}, one gets
\beq\label{cor}
\bra{\cal O}[g_{\mu\nu}]\ket^\tau=\int[\d\eta_{\mu\nu}]\,
{\cal O}[g_{\mu\nu}(\eta)]\exp\Big[\!-\demi\int \d^dx\d\tau\sqrt g\, 
\eta_{\alpha\beta}\,\NF^{-1}(G^{-1})^{\alpha\beta\rho\sigma}
\eta_{\rho\sigma}\Big].
\eeq
 
Each term in the Langevin equation \eqref{Lan} has a natural interpretation
as follows.
We will be especially concerned with their covariance.

The speed $D_\tau g_{\mu\nu}$ of $(\Sig,g_{\mu\nu})$ is the completion of the
viscous force $\pa_\tau g_{\mu\nu}$ by the term $2\nabla_{(\mu}N_{\nu)}$.
The Lie derivative in ${\cal M}$, as it is written in \eqref{Extri}, is a full
stochastic bulk operation.
However, one can define an intrinsic Lie derivative in $\Sig$,
$\Lie^{\Sig}_{{\pmb\xi}(x)}g_{\mu\nu}=2\nabla_{(\mu} \xi_{\nu)}$,
with a parameter equal to the $\Sig$-vector $\xi^\alpha(x)$. 
The term $2\nabla_{(\mu} g_{\nu)\tau }$ contained in $D_\tau g_{\mu\nu}$
enforces a drift force with parameter $N^\mu=N^2g^{\mu\tau}$ along the orbits
of $\Diff$ in the space of metrics $g_{\mu\nu}$.
It reproduces an infinitesimal action of $\Diff$ within $\Sig$ on the space
of metrics $g_{\mu\nu}$ with a parameter $N^\mu$, which expresses the
non-triviality of the foliation.
Thus, in the leaf's speed \eqref{speed} of the quantum field stochastic
evolution, the term $2\nabla_{(\mu}N_{\nu)}$ provides a gauge restoring
force along the gauge orbits of diffeomorphisms for observables that are
non-reparametrization invariant (as depending on unphysical longitudinal
degrees of freedom, which manifest themselves anyway in virtual processes). 
 
The term $2N g_{\mu\nu}$ reproduces a Weyl rescaling of the metric with a
parameter equal to the lapse $N$.
This gauge restoring force is needed for extracting Weyl dependent
observables, and the choice of $N$ will not affect the evolution of
Weyl-invariant ones. 
 
The lapse $N$ and shift vector $N^\mu$ are thus fields that play the role of
parameters for gauge-fixing restoring forces for unphysical degrees of freedom
in the Langevin stochastic evolution along orbits of the gauge symmetry
$\Diff\ltimes\Weyl$.

In fact, $N^\mu$ and $N$ are the gravitational analogue of the additional
gauge field component $A_\tau$ in the stochastic quantization of a gauge
field $A_\mu$, which gives a gauge fixing restoring force along the orbits
of Yang-Mills transformations and defines the Yang-Mills Parisi-Wu equation
in the complete space of gauge field configurations
\cite{Baulieu:1988jw, *Baulieu:1989qh, *Baulieu:1993ff, Baulieu:2000xi,BZ}.
The choices of $N^\mu$ and $N$ do not influence the $\tau=\infty$ limit of
the correlation functions of physical observables.\footnote{Appendix
\ref{App1} sketches a general proof that the first order stochastic evolution
of gauge invariant observables is not affected by additional gauge-fixing
restoring terms in the Langevin equation.
Any given choice of vector field $N^\mu$ gives the same evolution for an
observable that is reparametrization invariant.}
So, following the general strategy of
\cite{Baulieu:1988jw, *Baulieu:1989qh, *Baulieu:1993ff, Baulieu:2000xi},
we will perform functional integration over $N(x,\tau)$ and $N^\mu(x)$ in
the supersymmetric formulation of the Langevin equation, followed by some
BRST gauge-fixing on $N^\mu$.\footnote{The Weyl gauge restoring force term
$2Ng_{\mu\nu}$ in the Langevin equation \eqref{Lan} was not discussed in
\cite{Baulieu:2016qmc, *Baulieu:2018wju}.
We now understand that such a Weyl symmetry restoring force is necessary in
view of correctly defining observables.}

The way to obtain a supersymmetric representation of correlators
$\bra{\cal O}[g_{\mu\nu}]\ket^\tau$, where the noise have been integrated out,
is standard, as originally stated in \cite{Parisi:1980ys} and
\cite{1983PhLB..129..432G}.
Formally, it relies on determinant identities and the argument can be made
systematic in the context of topological quantum field theory, by imposing in
\eqref{cor} the Langevin equation \eqref{Lan} relating $\eta_{\mu\nu}$ and
$g_{\mu\nu}$ in a (stochastic) equivariant BRST invariant way.
This transforms \eqref{cor} into a supersymmetric path integral involving an
equivariant $Q$ supersymmetry acting on the field $g_{\mu\nu}$ and its
topological ghosts $\Ps_{\mu\nu}$ and $\bar\Ps_{\mu\nu}$.
One gets a $(d+1)$-dimensional TQFT path integral whose fields variables are
$g_{\mu\nu}$, $\Ps_{\mu\nu}$ and $\bar\Ps_{\mu\nu}$ with a $Q$-exact action
that localizes the path integral to the solution of the Langevin equation
\eqref{Lan}
\cite{Baulieu:1988jw, *Baulieu:1989qh, *Baulieu:1993ff, Baulieu:2000xi}.
The link between the Langevin equation and its supersymmetric representation
explains the choice that physical observables should be the restricted set of
functionals $\bra{\cal O}[\hat g_{\mu\nu}]\ket^\tau$, where $\hat g_{\mu\nu}$
is the unimodular part of $g_{\mu\nu}$ to be introduced shortly. 

\section{Irreducible decomposition of the Langevin equation and its kernel}\label{sec4}

We will irreducibly decompose each term of the Langevin equation \eqref{Lan}
in its traceless and trace parts, term by term. We need also to discuss the
positivity of the noise distribution, which is fundamental to guarantee
convergence.

The presence of the kernel $\NF G_{\mu\nu\rho\sigma}$ in front of Einstein
equations of motion in \eqref{Lan} is necessary.
Indeed, one must lower the indexes of the equations of motion appearing in
the Langevin equation to get a covariant equation.
Since $\NF G_{\mu\nu\rho\sigma}$ is a function of $g_{\mu\nu}$, $N^\mu$ and
$N$, the noise of \eqref{Lan} is currently named a multiplicative noise
in the language of statistical mechanics \cite{2009arXiv0901.1271K}.

The expression of the kernel reflects the gauge symmetry of the theory and
the tensor $G_{\mu\nu\rho\sigma}$ is fixed by requiring $\Diff$-covariance.
It is symmetric in $\mu\nu$ and $\rho\sigma$, so its general form is 
\beq\label{ker}
G_{\mu\nu\rho\sigma}(x,\tau)=\demi(g_{\mu\rho}g_{\nu\sigma}
+g_{\mu\sigma}g_{\nu\rho})-\lam g_{\mu\nu}g_{\rho\sigma},
\eeq
where $\lambda$ is a dimensionless constant. 
This tensor is nothing but a Wheeler-DeWitt metric
\cite{PhysRev.160.1113, *Giulini:1994dx} over the space of metrics
$g_{\mu\nu}$ of any given leaf in a linear space with dimension
$\frac{d(d+1)}{2}$.
In our case, we are free to choose the value of $\lam$, since we have the
kernel-independence theorem for the equilibrium distribution, valid for all
kernels that give a well-defined evolution.
This is in contrast to what happens in the different context when one uses a
Wheeler-DeWitt metric and the ADM formalism to tentatively quantize gravity
in the temporal gauge.
In this other situation, one gets a fixed value of the parameter $\lam$ by a
decomposition in $d-1$ dimensions of $d$-dimensional Einstein gravity, which
implies the damaging occurrence of a pseudo-Euclidean metric over the space
of $(d-1)$-dimensional spatial metrics.
In our case, the question is different, and for defining the stochastic
quantization of Euclidean gravity, we can use any value of $\lam$ that
ensures the positivity of the noise distribution.

The choice of the parameter $\lambda$ actually controls the Euclidean or
pseudo-Euclidean signature of the kernel \eqref{ker} through the sign of
$1-\lam d$.

Indeed the $\demi d(d+1)$ eigenvalues of $G_{\mu\nu\rho\sigma}$ are
\beq
(1-\lambda d,1,1,\dots,1).
\eeq
The positivity of $G_{\mu\nu\rho\sigma}$ is thus warranted if one chooses
the parameter $\lambda$ to satisfy 
\beq\label{cond>0}
\lam<\frac{1}{d}.
\eeq
Notice that also $\NF$ is guaranteed to be positive with our choice
\eqref{conds}.

For a more transparent discussion of the positivity of the noise weight, one
can decompose $G_{\mu\nu\rho\sigma}$ in traceless and trace parts.
Since $g^{\mu\nu}g^{\rho\sigma}G_{\mu\nu\rho\sigma}=d(1-d\lambda)$ one has
\beq
G_{\mu\nu\rho\sigma}=
G^T_{\mu\nu\rho\sigma}+\dfrac{1-d\lambda}{d}g_{\mu\nu}g_{\rho\sigma},
\eeq
with the traceless part
\beq
G^T_{\mu\nu\rho\sigma}=\demi(g_{\mu\rho}g_{\nu\sigma}
+g_{\mu\sigma}g_{\nu\rho})-\frac1d g_{\mu\nu}g_{\rho\sigma}.
\eeq
Denoting $\eta\equiv g^{\mu\nu}\eta_{\mu\nu}$, the noise 
$\eta_{\mu\nu}$ splits in traceless and trace parts:
\beq
\eta_{\mu\nu}=\eta^T_{\mu\nu}+\frac{1}{d}g_{\mu\nu}\eta.
\eeq
Defining the inverse $G^{-1}$ by
$(G^{-1})^{\mu\nu\rho\sigma}G_{\rho\sigma\alpha\beta}
=\delta_{(\alpha}^\mu\delta^\nu_{\beta)}$, one has
\beq
(G^{-1})^{\mu\nu\rho\sigma}=\demi(g^{\mu\rho}g^{\nu\sigma}+g^{\mu\sigma}
g^{\nu\rho})+\frac{\lambda}{1-\lambda d}g^{\mu\nu}g^{\rho\sigma}.
\eeq
Thus the definition of noise distribution \eqref{noise} is
\beq
\bra{\cal F}[\eta_{\alpha\beta}]\ket^\tau
=\int [\d\eta][\d\eta^T_{\alpha\beta}]\,{\cal F}[\eta,\eta^T_{\alpha\beta}]\,
\exp\Big[\!-\int\d^dx\d\tau\,\NF^{-1}\!\sqrt{g}
\Big(\eta^T_{\mu\nu}\eta^{T\mu\nu}+\frac{1}{d(1-\lambda d)}\eta^2\Big)\Big].
\eeq
This verifies that the Gaussian distribution has a positive definite weight
under the already spelled condition \eqref{cond>0}.

The gravity classical drift-force in the Langevin equation \eqref{Lan} is
the multiplication of the Einstein tensor
\be
E^{\rho\sigma}\equiv \frac{\delta S}{\delta g_{\rho\sigma}}
=R^{\rho\sigma}-\frac{1}{2} R g^{\rho\sigma}
\ee
by the positive kernel $\NF G_{\mu\nu\rho\sigma}$.
Its decomposition into the traceless and trace parts is
\bea
\NF G_{\mu\nu\rho\sigma}E^{\rho\sigma}
=\NF E^T_{\mu\nu}+\frac{(1-d\lambda)(2-d)}{2d}g_{\mu\nu}\,\NF R,
\eea
where $E^T_{\mu\nu}\equiv R_{\mu\nu}-\frac1d R g_{\mu\nu}$ is indeed traceless. 

Analogously, the stochastic speed decomposes into the traceless and trace
parts as
\beq
D_\tau g_{\mu\nu}\equiv D^T_\tau g_{\mu\nu} +\frac2d g_{\mu\nu}\k,
\eeq
where we defined
\bea
\k & \equiv & \pa_\tau\ln\sqrt{g}-\nabla_\mu N^\mu, \\
D^T_\tau g_{\mu\nu} & \equiv &
\pa_\tau g_{\mu\nu}-2\nabla_{(\mu} N_{\nu)}-\frac{2}{d}g_{\mu\nu}\k.
\eea

The stochastic acceleration ${\A}_{\mu\nu}$ decomposes as
\beq
\A_{\mu\nu}\equiv\A^T_{\mu\nu}+\frac1d\A g_{\mu\nu},
\eeq
where the trace and traceless parts are
\bea
\A&\equiv& g^{\mu\nu}{\A}_{\mu\nu}=g^{\mu\nu}(\pa_\tau-N^\alpha\pa_\alpha)
D_\tau g_{\mu\nu}-2g^{\mu\nu}D_\tau g_{\alpha\mu} \pa_\nu N^\alpha,  \\
\A^T_{\mu\nu}&\equiv& (\pa_\tau-N^\alpha\pa_\alpha) D_\tau g_{\mu\nu}
-2D_\tau g_{\alpha(\nu}\pa_{\mu)}N^\alpha-\frac1d g_{\mu\nu}\A.
\eea

By projecting onto the trace and traceless parts, we have finally decomposed
the Langevin equation \eqref{Lan} in irreducible representations with respect
to the internal diffeomorphism symmetry of the leaves:
\bea\label{LanTr}
\Del T\,\A+2\k&=&-\frac{\NF}{2}(1-d\lambda)(2-d)R+2dN+\sqrt\hbar\,\eta,\\
\Del T\,\A^T_{\mu\nu}+D^T_\tau g_{\mu\nu}&=&-\NF E^T_{\mu\nu}
+\sqrt\hbar\,\eta^T_{\mu\nu}.                       \label{LanTrless}
\eea
These irreducible decompositions play a central role in our analysis of the
Langevin equation.
We can already observe that there is no dependence on derivatives of the
lapse function $N$.

\section{Weyl transformation and unimodular decomposition}\label{sec5}

We just achieved the decomposition of the Langevin equation in irreducible
traceless and trace parts.

We now go a step further, by expressing these equations in function of the
unimodular part of the metric, the conformal factor and all other rescaled
Weyl-invariant fields. 
 
The change of field variables that extracts explicitly their Weyl
weight will illuminate some properties of the Langevin equation of gravity. 
In particular, it will allow us to decompose neatly the algebraic dependence
on $N$, with additional linear terms to the preexisting one in the gauge
restoring force $2Ng_{\mu\nu}$.
Using scale invariant fields turns out to be useful to define gravity
observables and reveal their Ward identities. 

A Weyl transformation of the metric is defined by
\beq 
\d s^2\mapsto\Omega^2 \d s^2\qquad\mbox{or}
\qquad g_{AB}\mapsto\Omega^2 g_{AB},
\eeq
where the Weyl factor $\Omega(x,\tau)$ is a function of all the coordinates.
Weyl transformations do not act on the coordinates but form a gauge symmetry. 
There is no practical need to introduce a Weyl gauge field for our purpose,
although it would make sense mathematically to do so. 

The way $\d s^2$ transforms implies the following Weyl transformation laws of
all metric field components of the $(d+1)$-dimensional ADM parametrization
\eqref{ADM}:
\beq
N\mapsto\Omega N, \qquad g_{\mu\nu}\mapsto\Omega^2 g_{\mu\nu},
\qquad N^\mu\mapsto N^\mu, \qquad \sqrt{g}\mapsto\Omega^d\sqrt{g}.
\eeq
The infinitesimal version of Weyl transformations, $\delta_\omega$, with
$\omega$ the infinitesimal abelian parameter, is
\beq
\delta_\omega N=\omega N, \qquad \delta_\omega g_{\mu\nu}=2\omega g_{\mu\nu},
\qquad\delta_\omega N^\mu=0\qquad\delta_\omega\frac{1}{d}\log\sqrt{g}=\omega.
\eeq
To extract Weyl-independent components in all fields, we use the decomposition
of the metric $g_{\mu\nu}$ in its unimodular part $\hat g_{\mu\nu}$ and
conformal factor $a\equiv \sqrt g^\frac1d$.\footnote{We checked that the
unimodular decomposition of all curvature tensors agree with
\cite{Kiefer:2017nmo}.
Notice however that our definition of $\hat K$ differs from the one in
\cite{Kiefer:2017nmo}, due to our ``golden rule''.}

The field $\phi\equiv\log a$ is in fact convenient to express the conformal
factor dependence.
One has
\beq 
\hat g_{\mu\nu}\equiv (\sqrt g)^{-\frac2d}g_{\mu\nu}, \qquad
a\equiv(\sqrt g)^\frac1d\equiv\exp\phi
\eeq
and hence
\beq
\delta_\omega\hat g_{\mu\nu}=0, \qquad \delta_\omega\phi=\omega.
\eeq 
As already stated, the shift vector $N^\mu$ is Weyl invariant from the
beginning.

The ADM fields of the $(d+1)$-dimensional space -- except $\phi$ -- can be
transformed into Weyl-invariant fields by appropriate rescaling:
\beq\label{weyl}
\ol N^\mu\equiv N^\mu, \quad \ol N \equiv a^{-1} N, \quad
\ol g_{\mu\nu}\equiv a^{-2} g_{\mu\nu},\quad
\ol g^{\mu\nu}\equiv a^{2} g^{\mu\nu}.
\eeq
All the curvatures and Lie derivatives involved in the Langevin equation of
gravity can be re-expressed using these fields.
Modulo a rescaling factor, they are made of terms involving only the hat
fields and of other terms involving derivatives of $\phi$. 

In what follows, every time an index $\mu$ is raised or lowered on a hatted
quantity, using the unimodular part of the metric, one gains a factor of
$a^2$ or $a^{-2}$ respectively (e.g. $\ol N_\mu\equiv a^{-2}N_\mu$).
To perform the unimodular decomposition of the Langevin trace and traceless
equations, one must proceed by successive steps.
Hatted quantities are defined by applying the following ``golden rule'':
for any given tensor $W$, $\hat W$ is defined from $W$ replacing all fields
by their hat rescaled ones.
In particular, since $\phi=\log a=\frac1d\log\sqrt{g}$, one has
$\hat\phi=\frac1d\log\sqrt{\hat g}=0$ because $\hat g_{\mu\nu}$ has unit
determinant.
In general, the application of this rule guarantees the Weyl-invariance of
the hatted quantities, since they are built out of Weyl-invariant objects
only.

A useful observation is that the Christoffel symbols decompose as 
$\Gam^\mu_{\nu\rho}=\ol\Gam^\mu_{\nu\rho}+\varSigma^\mu_{\nu\rho}$,
\cite{Thomas199, *Thomas2, *Thomas3}, 
with
\bea
\ol\Gam^\mu_{\nu\rho}&\equiv&\demi\ol g^{\mu\alpha}(\pa_\nu\ol g_{\alpha\rho}
+\pa_\rho \ol g_{\alpha\nu}-\pa_\alpha \ol g_{\nu\rho}),   \\
\varSigma^\mu_{\nu\rho}&\equiv& \delta^\mu_\rho\pa_\nu\phi+
\del^\mu_\nu\pa_\rho\phi-\ol g^{\mu\alpha}\ol g_{\nu\rho}\pa_\alpha\phi. 
\eea

Let us start with the unimodular decomposition of the various terms in
\eqref{LanTr} and \eqref{LanTrless}.
Consider first the Ricci tensor and curvature scalar of $g_{\mu\nu}$.
One has
\bea
R_{\mu\nu}&=&\hat R_{\mu\nu}-(d-2)\hat\nabla_\mu \pa_\nu \phi-\hat g_{\mu\nu}
\hat\nabla_\alpha \hat g^{\alpha\beta}\pa_\beta\phi+(d-2)\pa_\mu \phi\pa_\nu
\phi-(d-2)\hat g_{\mu\nu}\pa_\alpha\phi\hat g^{\alpha\beta}\pa_\beta\phi, \\
R&=&g_{\mu\nu}R^{\mu\nu}=\exp(-2\phi) \Big(\ol R-2(d-1)\ol g^{\mu\nu}\big(
\ol\nabla_\mu\pa_\nu\phi+\frac{d-2}{2}\pa _\mu\phi\pa_\nu\phi\big)\Big).
\eea
{}From now on, the ``hat $\Sig$-covariant derivative'' $\hat\nabla_\mu$ is
defined as $\nabla_\mu$ with the Christoffel symbols $\Gam^\mu_{\nu\rho}$
replaced by $\hat\Gam^\mu_{\nu\rho}$.
Likewise $\hat R_{\mu\nu}$ and $\hat R$ are obtained just like $R_{\mu\nu}$
and $R$ but using the hatted quantities $\hat\Gam^\mu_{\nu\rho}$ and
$\ol g_{\mu\nu}$. 
Defining
$\hat E^T_{\mu\nu}\equiv\hat R_{\mu\nu}-\frac1d\hat R\hat g_{\mu\nu}$,
one has
\bea\label{scale}
E^T_{\mu\nu}&=&\ol R_{\mu\nu}-\frac1d\ol g_{\mu\nu}\ol R
-(d-2)(\ol\nabla_\mu\pa_\nu \phi -\pa_\mu \phi\pa_\nu\phi)
+\frac{d-2}{d}\hat g_{\mu\nu}\hat g^{\alpha\beta}
\big(\hat\nabla_\alpha\pa_\beta\phi-\pa_\alpha\phi\pa_\beta\phi\big),  \nn\\
&=&\ol{E}^T_{\mu\nu}-(d-2)(\ol\nabla_\mu\pa_\nu\phi-\pa_\mu\phi\pa_\nu\phi)^T.
\eea
The scalar factor $\cal N$ decomposes as
\beq
\NF=\exp(2\phi)\,\hat\NF.
\eeq 
The trace of the stochastic speed is
\beq
\k=d(\pa_\tau-\hat N^\mu\pa_\mu)\phi-\hat\nabla_\mu \hat N^\mu
=d\tpa\phi-\hat\nabla_\mu \hat N^\mu, 
\eeq
where $\hat\nabla_\mu\hat N^\mu=\pa_\mu\hat N^\mu$ because
$\hat g_{\mu\nu}$ is unimodular and
\bea
\tpa\equiv\pa_\tau-\hat N^\mu\pa_\mu.
\eea
The traceless part of the stochastic speed is
\beq\label{hatTrl}
D^T_\tau g_{\mu\nu}=\exp(2\phi)\Big(\pa_\tau\hat g_{\mu\nu}
-2\hat\nabla_{(\mu} \hat N_{\nu)}+\frac2d \hat g_{\mu\nu}
\hat\nabla_\alpha \hat N^\alpha\Big)
\equiv\exp(2\phi)\hat D^T_\tau\hat g_{\mu\nu}.
\eeq
One has thus the following decomposition:
\beq
D_\tau g_{\mu\nu}=\exp(2\phi)\Big(\hat D_\tau\hat g_{\mu\nu}
+2\hat g_{\mu\nu}\tpa \phi\Big).
\eeq
The first term in the right hand side of this equation is an example of the
golden rule, with
\beq
\hat D_\tau\hat g_{\mu\nu}\equiv
\pa_\tau\hat g_{\mu\nu}-2\hat\nabla_{(\mu}\hat N_{\nu)}.
\eeq
Notice in particular that, while the trace term is not Weyl covariant due
to the $\phi$ dependence, the traceless part is Weyl covariant.
Stated differently, a Weyl transformation affects only the trace of the
stochastic speed.
 
Consider now the acceleration $\A_{\mu\nu}$ as it is defined in \eqref{acce}.
Its unimodular decomposition is
\beq
\A_{\mu\nu}=\exp(2\phi)\Big(\hat \gamma_{\mu\nu}+2\hat g_{\mu\nu}\tpa^2\phi
+4\hat g_{\mu\nu}(\tpa \phi)^2+4\hat D_\tau \hat g_{\mu\nu}\tpa \phi\Big).
\eeq
Here again the golden rule is at work: 
\beq
\hat\gamma_{\mu\nu}=\tpa\hat D_\tau \hat g_{\mu\nu}
-2\hat D_\tau\hat g_{\alpha(\mu}\pa_{\nu)}\hat N^\alpha.
\eeq
Consequently the trace part is
\beq
\gamma=\hat\gamma+2d \tpa^2\phi+4d(\tpa\phi)^2
-8\tpa\phi\hat\nabla_\mu \hat N^\mu,
\eeq
and the traceless part is
\beq
\gamma^T_{\mu\nu}=e^{2\phi}\Big(\hat\gamma^T_{\mu\nu}
+4\tpa\phi\hat D^T_\tau\hat g_{\mu\nu}\Big),
\eeq
where $\hat D^T_\tau\hat g_{\mu\nu}$ is defined in \eqref{hatTrl}.
Notice that $\hat\gamma$ and $\hat\gamma^T_{\mu\nu}$, which (in our knowledge)
are yet unknown quantities in the literature, are also given by the golden
rule.

One must also decompose using unimodular variables the noise, for which we
require
\beq
\eta_{\mu\nu}=e^{2\phi}\hat \eta_{\mu\nu}, \quad
\eta^T_{\mu\nu}=e^{2\phi}\hat\eta^T_{\mu\nu}, \quad \eta=\hat\eta.
\eeq

Putting everything together, one gets the following expression for the
traceless part of the second order Langevin equation \eqref{LanTrless}:
\beq\label{totoEtrl}
\Del T \Big(\hat\gamma^T_{\mu\nu}+4\tpa \phi \hat D^T_\tau\hat g_{\mu\nu}\Big)
+\hat D^T_\tau \hat g_{\mu\nu}=-\hat{\cal N}(\ol{E}^T_{\mu\nu}
-(d-2)(\ol\nabla_\mu\pa_\nu\phi-\pa_\mu\phi\pa_\nu\phi)^T)
+\sqrt\hbar\,\hat\eta^T_{\mu\nu}.
\eeq
As a result of our choice for the kernel, this equation is
homogeneous in $\phi$ with an overall linear dependence on $\hat\NF$. 

The trace part of \eqref{LanTrless} is: 
\bea\label{totoE}
&\Del T (\hat\gamma+2d \tpa^2\phi+4d(\tpa\phi)^2-8\tpa\phi\hat\nabla_\mu
\hat N^\mu)+2(d\tpa\phi-\hat\nabla_\mu\hat N^\mu)         \nn      \\
&=\frac{\hat{\cal N}}{2}(1-d\lambda)(d-2)\Big(\ol R-2(d-1)\ol g^{\mu\nu}
\big(\ol\nabla_\mu\pa_\nu\phi+\frac{d-2}{2}\pa _\mu\phi\pa_\nu\phi\big)\Big)
+2d\exp(\phi)\hat N+\sqrt\hbar\,\hat\eta.
\eea

In $d=2$ the traceless part does not depend on $\phi$, and so
are the observables we are looking for.
In this case the use of the unimodular part $\hat g_{\mu\nu}$ amounts 
to that of the Beltrami differential.
See the comment just below.

For $d>2$, the traceless parts depends on $\phi$, but it is homogeneous,
a property that we shall shortly interpret. 

For completion, by using the unimodular parametrization of the noise, one
gets the noise distribution in the following form:
\be\label{totoN}
\bra{\cal{F[\hat\eta_{\mu\nu}]}}\ket^\tau=\int[\d\hat\eta]
[\d\hat\eta^T_{\mu\nu}]\,{\cal F}[\hat\eta,\hat\eta^T_{\mu\nu}]
\exp\Big[\!-\demi\int\d^dx\d \tau \exp(-2\phi)\hat\NF^{-1}
\big(\hat\eta^T_{\alpha\beta}\hat\eta^{T\alpha\beta}
+\frac{1}{1-\lambda d}\hat\eta^2\big)\Big].
\ee 

The result of \cite{York72, *York73} that the classical evolution is
determined by the conformal classes of metric $g_{\mu\nu}$ suggests that
we may consider $\hat g_{\mu\nu}$ and $\phi$ as the independent field
variables of gravity, where only $\hat g_{\mu\nu}$ is physical.
If correct, and we will show it in the first order Langevin theory, this
would mean that the stochastic time evolution of the conformal factor is
basically irrelevant for physical observables, which we claim being Weyl
invariant. 

We remark that defining observables as only depending on the
unimodular part of $g_{\mu\nu}$ ensures from the beginning that the physical  quantum graviton, seen as an excitation of  $g_{\mu\nu}$, must be traceless and symmetric. This traceless property  is obvious in classical and semi-classical gravity, but it  must rely on some symmetry principle in any given  attempt of defining quantum gravity. Our definition of quantum observables is compatible  with the  classical property.

What has been done above for $d>2$ is quite a striking generalization of the
case $d=2$, \cite{Baulieu:1986hw, *Baulieu:1987jz, *Baulieu:1989dg}, where
one can use efficiently the Beltrami differential $\mu\equiv\mu^z{}_{\bar z}$
as the only fundamental field variable of the $2$-dimensional metric, with
$\hat g_{\mu\nu}=\frac1{1-\mu\bar\mu}{1\quad\mu\choose\bar\mu\quad1}$. 
Despite of the fact that in $2$ dimensions $\hat g_{\mu\nu}$ does not
propagate but intervenes only through its constant moduli, the (complex)
Beltrami $1$-form $\d z+\mu\d\bar z$ is actually the physical field in
$2$-dimensional gravity because the conformal factor is decoupled, and
possibly substituted with an additional Liouville field in non critical
dimensions.
If needed, the latter is seen as a Wess-Zumino field within the context
of a conformal theory, for one has to compensate the conformal anomaly
\cite{Polyakov:1981re, *Polyakov:1981rd, Baulieu:1986hw, *Baulieu:1987jz,
*Baulieu:1989dg}.
For $d>2$, one has of course the very non-trivial
propagation of $\hat g_{\mu\nu}$ in contrast to the case $d=2$, but,
nonetheless, the definition of observables is analogous. 

\section{Observables} \label{sec6}

{}From now on we will focus on the choice that $\NF$ satisfies \eqref{conds},
i.e.,
\beq
\NF=N^\alpha N_\alpha.
\eeq
This will allow us solve algebraically for $\hat N$ in the regime $\Del T=0$
and obtain a consistent definition of the expectation value of observables
${\cal O}[\ol g_{\mu\nu}]$ so that it depends only on the unimodular part of
the metric.
The analysis is simpler when $\Del T=0$, a situation that we now detail.

\subsection{$\pmb{\Del T=0}$}

Consider the Langevin equation \eqref{totoEtrl} and \eqref{totoE} with
$\Del T=0$.
In the supersymmetric formulation, the path integral of correlators of the
unimodular part of the metric is
\bea\label{Lansusy}
&&\bra {\cal O}[\ol g_{\mu\nu}] \ket^\tau=\int[\d\hat N][\d\hat N^\mu]
[\d\hat g_{\mu\nu}][\d\phi]\,{\cal O}[\ol g_{\mu\nu}]\,\exp\Big[\!
-\frac1{2\hbar}\int\d^dx\d\tau\frac{\exp(-2\phi)}{\hat N^\alpha\hat N_\alpha}
                                                                  \nn\\
&&\big(\|\hat D^T_\tau\hat g_{\rho\sigma}+\hat N^\alpha\hat N_\alpha
(\ol{E}^T_{\rho\sigma}-(d-2)(\ol\nabla_\rho\pa_\sigma\phi-\pa_\rho\phi
\pa_\sigma\phi)^T)\|^2
+\frac{1}{1-\lambda d}\Big\|2(d\tpa\phi-\hat\nabla_\beta\hat N^\beta)-    \nn\\
&&-\frac{\hat N^\alpha\hat N_\alpha}{2}(1-d\lambda)(d-2)
\Big(\ol R-2(d-1)\big(\ol\nabla^\beta\pa_\beta\phi+\frac{d-2}{2}
\pa^\beta\phi\pa_\beta\phi\big)\Big)-2d \exp(\phi)\hat N\Big\|^2+
\mathrm{susy \ terms}\big)\Big].                                                       \nn\\
\eea

We have only made explicit the bosonic part of the $(d+1)$-dimensional action.
The part $susy\ terms$ contains fermionic terms that form a fermionic path
integral that takes care of the Jacobian of the map between the fields and
the noise, which ensures the stochastic supersymmetry.
See e.g. \cite{1983PhLB..129..432G, Baulieu:1988jw, *Baulieu:1989qh,
*Baulieu:1993ff, Baulieu:2000xi} for details.\footnote{The total action,
including fermionic terms, is $Q$ exact under the topological stochastic
BRST operator
$Q\hat g_{\mu\nu}=\hat\Ps_{\mu\nu}+\Lie^{\Sig}_\xi\hat g_{\mu\nu}$,
$Q\xi^\mu=-\Ph^\mu+\xi^\nu\pa_\nu\xi^\mu$,
$Q\hat\Ph^\mu=\xi^\nu\pa_\nu\Ph^\mu-\Ph^\nu\pa_\nu\xi^\mu $,
following the same pattern as the stochastic quantization of the Yang-Mills
case. 
The details of the by-now standard method is not worth being displayed here,
since we only want to come back to the Langevin equation after the decoupling
of the conformal factor by a consistent gauge-fixing.
In fact this decoupling preserves $Q$ symmetry, so that we can go back to
the Langevin equation after showing how it works in the bosonic sector only.}
There is no need here to make these terms explicit.
 
Because ${\cal O}[\ol g_{\mu\nu}]$ is Weyl invariant, this path integral is
invariant under the Weyl transformation defined in \eqref{weyl}, but it
depends on the parameter $\tau$, as readily seen in \eqref{Lansusy}.

This $x$-independent Weyl symmetry of the path integral representation of
observables $\bra{\cal O}[\ol g_{\mu\nu}]\ket^\tau$ is nothing but a
dilatation of the lapse function $N$ by the same factor everywhere in any
given leaf.
It is a symmetry under any reparametrization of the stochastic time
$\tau\to\tau'(\tau)$, which is a sophisticated generalization of the
worldline reparametrization invariance of the relativistic particle theory.

The invariance of the path integral \eqref{Lansusy} can be indeed explained
by a BRST symmetry operation~$s$ using an abelian anticommuting ghost
$\omega(\tau)$, which completes the anticommuting vector ghost $\xi^\mu(x)$
accounting for diffeomorphisms in each leaf.
The operator $s$ acts on all fields as a nilpotent graded differential
operation with 
\bea\label{brstw}
s\ol g_{\mu\nu}&=&\Lie^\Sig_\xi\ol g_{\mu\nu} 	  \nn \\
s\phi&=&\Lie^\Sig_\xi\phi+\omega (\tau) 
\nn \\
s\hat N^\mu&=&\Lie^\Sig_\xi\hat N^\mu=\xi^\nu\pa_\nu\hat N^\mu
-\hat N^\nu\pa_\nu\xi^\mu.
\eea 
The condition $s^2=0$ on $\ol g_{\mu\nu}$ and $\phi$ implies the following
transformations of the ghosts
\bea
s\xi^\mu(x)&=&\xi^\nu\pa_\nu\xi^\mu, \nn \\
s\omega(\tau)&=&0.
\eea
The $s$ invariance of the action is achieved provided $\hat N$ transforms as
\beq\label{refN}
s\hat N=s\Big(\exp(-\phi)(\tpa\phi-\frac1d\hat\nabla_\beta\hat N^\beta)
-\frac{\exp(-\phi)\hat N^\alpha\hat N_\alpha}{4d}(1-d\lambda)(d-2)
\Big(\ol R-2(d-1)\big(\ol\nabla^\beta\pa_\beta\phi
+\frac{d-2}{2}\pa^\beta\phi\pa_\beta\phi\big)\Big)\Big),
\eeq
Since $s\hat N$ is an $s$-exact expression of $\hat g_{\mu\nu}$,
$\hat N^\mu$, $\phi$, $\xi^\mu $ and $\omega$ and since $s^2=0$ on these
fields, one has eventually $s^2\hat N=0$ on all fields.
We will shortly eliminate $\hat N$ by its equation of motion, so there is
no need to write $s\hat N$ explicitly.

Using the $x$-independent Weyl invariance and the BRST symmetry we just
defined, one can localize $\phi$ in an $s$-invariant way to the gauge choice
\bea\label{phig}
\phi(x,\tau)=\phi(x)\equiv\phi_{\{x\}}.
\eea
We thus reach the conclusion that the conformal factor is a spectator in
the stochastic evolution of observables, which is just given by some initial
condition.
More precisely, we found that, for observables
$\bra{\cal O}[\ol g_{\mu\nu}]\ket^\tau$, we can use a conformal factor
$\phi_{\{x\}}$ that is independent of $\tau$ in a BRST invariant way.
Therefore in the $\tau$ evolution it remains equal to some arbitrarily chosen
initial data $\phi_{\{x\}}$ which is thus a stochastic time independent
background for the evolution of observables. In the classical limit, which
is the only possible limit when $\tau=\infty$, it becomes the standard
conformal factor in general relativity.

Notice furthermore that $\hat N$ has a $\tau$ evolution that is basically
governed by the internal and external conformal scalar curvatures of the leaf
$\hat R(\hat g_{\mu\nu}(x,\tau))$ and $\hat K(\hat g_{\mu\nu}(x,\tau))$,
according to its equation of motion
\beq
\hat N=\exp(-\phi)(\tpa\phi-\frac1d\hat\nabla_\beta\hat N^\beta)
-\frac{\exp(-\phi)\hat N^\alpha\hat N_\alpha}{4d}(1-d\lambda)(d-2)
\Big(\ol R-2(d-1)\big(\ol\nabla^\beta\pa_\beta\phi+\frac{d-2}{2}
\pa^\beta\phi\pa_\beta\phi\big)\Big).
\eeq

For unphysical correlators, which are $\phi$ dependent, one can also proceed
to the elimination of $\hat N$, but one obtains a path integral that is not
$s$-invariant: $\phi$ is no more a spectator and has a $\tau$-evolution.

After the elimination of the lapse function $\ol N$ by its algebraic equation
of motion, the path integral \eqref{Lansusy} reads (skipping the ghost term
dependence that ensures Weyl BRST invariance) 
\bea\label{Lansusyconf}
&&\bra {\cal O}[\ol g_{\mu\nu}]\ket^\tau=\int[\d{\hat N}^\mu]
[\d\hat g_{\mu\nu}][\d\phi_{\{x\}}]\,{\cal O}[\ol g_{\mu\nu}]\,
\exp\Big[\!-\frac1{2\hbar}\int\d^dx\d\tau
\frac{\exp(-2\phi_{\{x\}})}{\hat N^\alpha\hat N_\alpha}            \nn\\
&&\big(\|\hat D^T_\tau \hat g_{\rho\sigma}+\hat N^\alpha\hat N_\alpha
(\ol{E}^T_{\rho\sigma}-(d-2)(\ol\nabla_\rho\pa_\sigma\phi_{\{x\}}
-\pa_\rho \phi_{\{x\}}\pa_\sigma\phi_{\{x\}})^T)\|^2
+\mathrm{susy\ terms}\big)\Big].
\eea
This is eventually the genuine path integral definition for physical
observables. 

Because one functionally integrates over all possible $\ol N^\mu(x)$, and
because the action is invariant under diffeomorphisms in each leaf, one must
do a BRST invariant gauge fixing on $\ol N^\mu(x)$ to get a propagation with
no zero modes in $\hat g_{\mu\nu}$.
Indeed, exploiting the reparametrization invariance of the observables and
the action, we can use for example the gauge fixing choice 
$\hat N^\mu(x)=\pa_\nu\hat g^{\mu\nu}(x,\tau_0)$. 
This gauge fixing is a good candidate for fixing the internal
reparametrization gauge in \eqref{Lansusyconf}, as can be seen for instance
by doing an expansion of $\hat g_{\mu\nu}$ around a classical background
with a small excitation. 
To regularize potential singularities in space of $\hat g_{\mu\nu}(x,\tau_0)$,
one may in fact add to the gauge fixing function a nowhere vanishing constant
vector $n^\mu$ chosen in the initial leaf, that is, 
\be\label{dif}
\hat N^\mu(x)=\pa_\nu\hat g^{\mu\nu}(x,\tau_0)+n^\mu.
\ee

Having obtained \eqref{Lansusyconf} and defined the gauge choice \eqref {dif},
we can come back to the Langevin equation and eliminate the $susy\ terms$ by
doing the usual change of variables that connects the Langevin equation to
its supersymmetric representation.
This equation, which only depends on a transverse noise has the form
\beq\label{gl}
\hat D^T_\tau \hat g_{\rho\sigma}
=-\hat N^\alpha\hat N_\alpha(\ol{E}^T_{\rho\sigma}
-(d-2)(\ol\nabla_\rho\pa_\sigma\phi_{\{x\}}
-\pa_\rho\phi_{\{x\}}\pa_\sigma\phi_{\{x\}})^T+\sqrt\hbar\,\eta^T _{\mu\nu},
\eeq
with a positive $\hat g_{\mu\nu}$-Gaussian-norm for the noise
$\eta^T_{\mu\nu}$.
In \eqref{gl}, the $\tau$-Weyl symmetry is manifest and $\phi_{\{x\}}$ is
just a spectator.
It could be called the Parisi--Wu equation of gravity.
The regime with $\Del T=0$ should hold near the end of the transition where
gravity becomes classical.
The Langevin equation \eqref{gl} could have been postulated from the
beginning, for it is a consistent equation, if one postulates the
$\tau$-Weyl symmetry of observables. 
Nonetheless, it is a rewarding fact that it has been extracted from the
geometrical equation \eqref{Lan}, involving all ingredient of the foliated
$(d+1)$-dimensional space. 

A non-trivial and interesting feature that has been developed in this section
is that for computing physical observables, one can fix the lapse and shift
of the foliation \eqref{refN} and \eqref{dif} in function of $\phi_{\{x\}}$
and $\hat g_{\mu\nu}(x,\tau_0)$.
The conformal factor $\phi_{\{x\}}=\phi(x,\tau_0)$ is a spectator in the
stochastic process, fixed by its initial condition.
In contrast, the lapse $N(x,\tau)$ has a $\tau$-evolution, determined by both
the extrinsic and intrinsic scalar curvatures of each leaf.

\subsection{$\pmb{\Del T\neq0}$}
 
This situation has a profound difference with respect to the previous one,
and it is the one that was heuristically predicted in
\cite{Baulieu:2016qmc, *Baulieu:2018wju}, because of the acceleration term
that provides oscillations of functions of $\tau$, at a scale of $\Del T$,
giving creations and annihilation of quanta in the $(d+1)$-dimensional theory.

Indeed, in the acceleration term of the traceless part of the Langevin
equation, there is a term proportional to $\pa_\tau\phi$:
\beq
\Del T \Big(\hat\gamma^T_{\mu\nu}+4\tpa\phi\hat D^T_\tau\hat g_{\mu\nu}\Big)
+\hat D^T_\tau \hat g_{\mu\nu}=-\hat{\cal N}( \ol{E}^T_{\mu\nu}
-(d-2)(\ol\nabla_\mu\pa_\nu\phi-\pa_\mu \phi\pa_\nu\phi)^T)
+\sqrt\hbar\,\hat\eta^T_{\mu\nu}.
\eeq

The additional term
$\Del T(\hat\gamma^T_{\mu\nu}+4\tpa\phi\hat D^T_\tau\hat g_{\mu\nu})$
brings new insightful physics.
Two main regimes can be distinguished.
\begin{enumerate}
\item The regime
$\hat\gamma^T_{\mu\nu}\gg4\tpa\phi\hat D^T_\tau\hat g_{\mu\nu}$.
In this case, $\phi$ is still non dynamical, as it is for the first order
equation.
Consequently it can be gauge fixed again as in \eqref{phig}.
This scenario, although technically harder and conceptually different due to
the second order term, can be treated using the same analysis as for
$\Del T=0$.
That is, one can solve algebraically the trace part for $\hat N$, inject the
result in the traceless part and gauge fix $\phi$.
The result is a traceless second-order Langevin equation for
$\hat g_{\mu\nu}$.
Notice however that there are oscillations here due to the second order term,
such that an equilibrium distribution is not reachable.
This indicates that this regime is probing deeper in the stochastic bulk, but
it is an intermediate step in the stochastic evolution, because the conformal
factor is still non-dynamical.
Mathematically this is encoded in the fact that, although the leaves are
oscillating, the conformal factor is dictated by its initial value.
To further analyze this case, we report explicitly its traceless Langevin
equation in the simplified situation where $N^\alpha$ and $\phi_{\{x\}}$
are constant:
\beq
\Del T (\tilde\pa^2_\tau\hat g_{\mu\nu})^T+\tilde\pa_\tau \hat g_{\mu\nu}
=-\hat{\cal N}\Big(\hat R_{\mu\nu}-\frac1d \hat R\hat g_{\mu\nu}\Big)
+\sqrt\hbar\,\hat\eta^T_{\mu\nu}.
\eeq
This situation is thus a typical second order Langevin theory,
\cite{Baulieu:2016qmc, *Baulieu:2018wju}, but for the unimodular part of
the metric.

\item The regime
$\hat\gamma^T_{\mu\nu}\ll4\tpa\phi\hat D^T_\tau\hat g_{\mu\nu}$.
This means that, deep in the stochastic bulk, the Weyl part of the symmetry
\eqref{brstw}, allowing to set \eqref{phig}, is lost.
Therefore the stochastic temporal evolution of the unimodular part of the
metric is influenced by the evolution of the conformal factor, which is no
more a spectator.
We must thus compute the evolution of observables doing the path integral
over the field $\phi(x,\tau)$ as well.
This is a non-trivial task, although it is a well-defined problem.
The idea is the following: the stochastic evolution starts with some initial
configuration at a fixed stochastic time $\tau_0$ in the bulk.
Stochastic second order quantum effects make the temporal evolution of
$\hat g_{\mu\nu}(\tau_0,x)$ and $\phi(\tau_0,x)$ rapidly oscillate.
At a late time, an abrupt transition brings the full Langevin equation to
the first order one.
There, the evolution of $\hat g_{\mu\nu}$ does not depend on the
one of $\phi$, and we retrieve the physics discussed in the $\Del T=0$ case,
with a well-defined equilibrium distribution at late stochastic time.
The simplified case with $N^\alpha$ and $\phi_{\{x\}}$ constant now reads
\beq
\tilde\pa_\tau \hat g_{\mu\nu}(1+4\Del T \tilde\pa_\tau \phi)
=-\hat{\cal N}\Big(\hat R_{\mu\nu}-\frac1d \hat R\hat g_{\mu\nu}\Big)
+\sqrt\hbar\,\hat\eta^T_{\mu\nu}.
\eeq
It is clear that, as long as $\Del T\neq 0$, there is a coupling between the
evolution of $\hat g_{\mu\nu}$ and $\phi$.
In other words, the stochastic evolution is not unimodular in the deep bulk.
Here observables are therefore functionals of $\phi$ and the discussions of
the previous section are no longer applicable. This regime constitutes an
appealing direction of investigation, with its complete understanding yet
to unravel.
\end{enumerate}

In both cases, one must separates the regime where the acceleration term
dominates the friction and vice-versa. In the former the theory is dominated
by oscillations, where there cannot be an equilibrium, i.e., the theory
remains $(d+1)$-dimensional with no possibility of defining a Lorentz time
in each leaf.
See \cite{Baulieu:2016qmc, *Baulieu:2018wju} for some heuristic description
of the resulting physics for $\Del T\neq 0$.
The latter is instead the situation treated in the previous section, where it
is certain that an equilibrium solution at late time exists.
This regime can stop when there is a fluctuation where effectively gravity
becomes classical, so one can neglect the noise because it is factorized by
$\sqrt\hbar$, in which case one quickly reaches the equilibrium at large
values of $\tau$, such that the theory can be computed in the bulk, with the
possibility of using a Lorentz time, just by solving classical equations of
motion.
This, in its cosmological application, means that we passed the phase
transition marked by the inflation.

\section{Conclusion}\label{sec7}
 
In the series of work \cite{York72, *York73} it was clearly mentioned that
what matters when solving the classical Einstein equations of motion is the
propagation of conformal classes of spatial metrics.
In fact the issue of giving a role to Weyl symmetry for the Einstein theory
can be traced back to a time as remote as 1925
\cite{Thomas199, *Thomas2, *Thomas3}.
In order to be consistent with the well-established property that the
equations of motion of classical gravity make no relevant difference between
metrics in the conformal class, although the gravity action is not Weyl
invariant, we raised as a principle the definition of quantum gravity
observables of the metrics as covariant functionals of their
unimodular parts.
 
In fact, beyond technical difficulties, the definition of the observables of
a theory is a notion that goes prior to the choice of the method that one
chooses to quantize a classical theory. So, we pointed out that the gauge
symmetry that determines the observables of gravity is generally Weyl
invariance.

To render this explicit, we proposed to use stochastic quantization for
defining quantum gravity, a feature that we originally introduced with the
motivation that it allows to bypass the question of the impossibility of
defining the Lorentz time when quantum gravity is switched on.
We have proposed a seemingly consistent way to define the observables of
Euclidean quantum gravity. 
 
We have shown that the various properties of stochastic quantization define
a process where Weyl symmetry is maintained for physical observables along
their propagation in stochastic time.
The theory predicts the stochastic lapse and shift being determined as a
function of initial conditions.
Ward identities imply that the conformal factor is physically irrelevant in
quantum gravity, at least at late stochastic time.
Specifically, this holds for the first order Langevin equation and the second
order in the regime where the $\hat g_{\mu\nu}$ fluctuations are greater than
$\pa_\tau\phi$.
We found also another second order regime, opposite to the one just depicted,
where Weyl symmetry is absent due to temporal evolution of $\phi$.
This scenario is not treatable with the analysis of this paper.
It deserves further study.
 
The physical irrelevance of the conformal factor in first order stochastic
quantum gravity is an unexpected and pleasant generalization of the soluble
case of two dimensions.
We now believe that Weyl invariance should be postulated as the gauge symmetry
principle of gravity in general.
On the other hand reparametrization invariance is an internal freedom of the
theory encoding the fact that one can choose mathematically any (consistent) 
set of coordinates in each leaf of a foliation of the $(d+1)$-dimensional
space $\{x,\tau\}$, determined by the stochastic time evolution. 
 
One cannot exclude the possibility that a conformal anomaly can occur.
If it is the case, there will be no conceptual difficulty to establish the
conformal invariance of quantum gravity by introducing a Wess--Zumino field,
following basically the same pattern as
\cite{Polyakov:1981re, *Polyakov:1981rd}, where the Liouville field for
non-critical strings in the case $d=2$ was introduced.

We conclude with a remark concerning the geometrical setup of this work.
Having a $d$-dimensional theory connected to a $(d+1)$-dimensional one is
a feature shared by many modern constructions.
For instance this happens in holography, where one defines a gravitational
theory in $d+1$ dimensions which is related to a $d$-dimensional matter
theory living on its boundary.
Although here we use an extra dimension to discuss the stochastic enhancement
of the boundary theory, we have a stochastic flow analogous to a holographic
RG flow. 

\appendix

\section{On the gauge fixing restoring forces}\label{App1}

In this Appendix we sketch a proof that for the first order Langevin equation
additional gauge fixing restoring terms do not modify the evolution of gauge
invariant observables, as it was introduced for the Yang-Mills theory in
\cite{BZ}.

Consider the generalization of the first order Langevin equation
$\alpha\dot q=-U'(q)+\beta\eta$ when we have a quantum field theory with
a gauge symmetry.
Replace $q$ by a $\fp(x,\tau)$ and $U(q)$ by its action $I[\fp]$.
Let $\delta^\gauge_\epsilon(x,\tau)$ be the gauge transformation of $\fp$
with a local parameter $\epsilon$.
The gauge invariance of the action means $\delta^\gauge_\epsilon(I)=0$ and
a gauge invariant observable is a functional ${\cal{O }}_{G-I}$ with 
\bea
\int\d\tau\d x\,\delta ^\gauge_\epsilon(\fp(x,\tau))
\frac{\delta{\cal{O}}_{G-I}}{\delta \fp(x,\tau)}=0.
\eea
The generalization of the Langevin equation with a kernel $K$ and a gauge
restoring force along the orbits of the gauge transformation depending on
an arbitrarily chosen functional $v(\fp)$ is
\bea\label{gag}
\frac{\pa\fp}{\pa\tau}=K\Big(\frac{\delta I}{\delta\fp}+\delta ^\gauge_v(\fp)
\Big)+\eta,
\eea
with the probability distribution for the noise
\bea \bra{\cal{F } } [\eta(x,\tau)]\ket^\tau=
\int [\d \eta]_{x,\tau}\,{\cal F}[\eta(x,\tau)]\,
\exp\Big[\!-\int\d\tau'\d x'\eta(x',\tau')K^{-1}\eta(x',\tau')\Big].
\eea
The products by $K$ of $\frac{\delta I}{\delta \fp} $ and
$\delta^\gauge_v(\fp)$ are respectively drift forces along the physical
excitations of $\fp$ and unphysical gauge excitations of $\fp$, respectively.
$\delta^\gauge_v(\fp)$ can be called a stochastic gauge fixing force with
field parameter $v$.

Eq.~\eqref{gag} implies a Fokker-Planck equation that computes
equal-stochastic-time correlators, with
\bea
\bra{\cal{O } } [\fp(x,\tau)]\ket^{\tau, v} = 
\int [\d \fp]_y\,P^v (\fp_y,\tau)\,{\cal{O}}[\fp_y]
\eea
and 
\bea\label{FPl}
\frac{\pa P^v (\fp,\tau)}{\pa \tau}=\Big[\int\d y\,\frac{\delta}{\delta\fp_y}
K\Big(\frac{\delta}{\delta\fp_y}+\frac{\delta I}{\delta\fp_y}
+\delta^\gauge_v(\fp_y)\Big)\Big]P^v(\fp,\tau).
 \eea
 
If $v$ is a local function of $\fp$, the equilibrium distribution is defined
when $\tau\to\infty$ and depends on $v$ in a non-local way.
Equation \eqref{FPl} shows also that if there is a normalizable stationary
distribution $P^v(\fp,\tau=\infty)$ for the equilibrium distribution
that is reached smoothly, it is independent on $K$. 

On the other hand, if one performs a supersymmetric representation of
the Langevin equation \eqref{gag}, locality can be enforced by functionally
integrating over all possible choices of $v$, introduced as an independent
field.
One must then proceed to a BRST-invariant gauge fixing of $v$, a task that
can be done in a way that is compatible with stochastic supersymmetry. 
 
Now we can compute $\bra\frac{\pa{\cal O}[\fp(x,\tau)]}{\pa\tau}\ket^\tau$
using the Fokker-Planck equation \eqref{FPl}.
After an integration by parts one gets
\bea
\Big\langle\!\!\Big\langle\frac{\pa{\cal O}[\fp(x,\tau)]}{\pa\tau}
\Big\rangle\!\!\Big\rangle^\tau
=\int[\d\fp]\,\Big[\int\d y\,K\Big(\frac{\delta}{\delta\fp_y}
-\frac{\delta I}{\delta\fp_y}-\delta^\gauge_v(\fp_y)\Big)
\frac{\delta}{\delta \fp_y}\Big]{\cal O}[\fp(x,\tau)].
\eea
Since on the right hand side
$\int\d y\,\delta^\gauge_v(\fp_y))\frac{\delta}{\delta\fp_y}$ acts as a gauge
transformation with parameter $v$ on functionals of $\fp$, we see that if
${\cal O}$ is a gauge invariant functional, the last term cancels and the
evolution $\bra\frac{\pa{\cal O}[\fp(x,\tau)]}{\pa\tau}\ket^\tau$ of
${\cal O}[\fp(x,\tau)]$ is independent on $v$.
 
On the contrary, the evolution of non gauge invariant observables depends on
the choice of $v$, whose presence is actually necessary in order to define
the evolution itself.
 
To compute both gauge-invariant and non-gauge-invariant correlators, one
either defines a clever choice of $v$ or considers $v$ as an independent
field and integrates over all possibilities with a BRST-invariant gauge
fixing of $v$.

Both strategies are legitimate, provided the choice of function $v$ or gauge
fixing gives a well defined result.
 
As always, there are good choices of gauge versus bad choices.
One expects good classes of gauges governed by some parameters.
For instance in the Yang-Mills case, the class of gauges
$v\equiv A_5 =\alpha\pa_\mu A_\mu$ determines a perfectly well defined
stochastic gauge-fixing, with $\alpha$ a free parameter.
In this case, one can in fact prove rigorously in perturbation theory that
physical correlators are $\alpha$-independent \cite{Baulieu:2000xi}. 

In this paper devoted to gravity, we did the gauge choice of
Eqs.~\eqref{phig} and \eqref{dif}. 

When we have acceleration, the dependence on $v$ is more subtle.
However, when the evolution is dominated by the friction and we are near
the equilibrium, the theorem applies.

\providecommand{\href}[2]{#2}
\begingroup\raggedright\endgroup

\end{document}